 \documentclass[11pt]{article}
\newtheorem{defn}{Definition} 

\newtheorem{thm}{Theorem}
\newtheorem{lem}[defn]{Lemma}
\newtheorem{cor}{Corollary}
\newtheorem {rem}{Remark}
\newtheorem{con}{Conjecture}

\title{\v{C}erny-Starke conjecture from the sixties of  XX century}

\date{2021}

\author{A.N. Trahtman}

\begin{document}

\maketitle

  \begin{abstract}
  A word $s$ of letters on edges of underlying graph $\Gamma$ of deterministic finite automaton (DFA)
is called synchronizing if $s$ sends all states of the automaton to a unique state.

J. \v{C}erny discovered in 1964 a sequence of $n$-state complete DFA possessing a minimal synchronizing word
of length $(n-1)^2$.
The hypothesis, mostly known today as \v{C}erny conjecture, claims that $(n-1)^2$ is a precise upper bound on
the length of such a word over alphabet $\Sigma$ of letters on edges of $\Gamma$ for every complete $n$-state DFA.
The hypothesis was formulated in 1966 by Starke.

To prove the conjecture, an algebra with non-standard operations on a special class of matrices, induced by words in the alphabet of labels on edges, is used. These matrices with respect to the aforementioned operation form a space with a zero matrix as a neutral element.

The proof is based on the connection between lengths of words
$u$ and dimension of the space generated by solution $L_x$ of matrix equation $M_uL_x=M_s$ for synchronizing word $s$,
as well as on relation between ranks of $M_u$ and $L_x$.
Important role below placed the notion of pseudoinvers
matrix.
\end{abstract}

$\bf Keywords:$ deterministic finite automata, minimal synchronizing word, \v{C}erny conjecture

\section*{Introduction}

 The problem of synchronization of DFA is a natural one and various aspects of this problem have been touched in the literature.
Synchronization makes the behavior of an automaton resistant
against input errors since, after detection of an error, a
synchronizing word can reset the automaton back to its original
state, as if no error had occurred. An occurrence of a synchronizing word limits the propagation of errors for a prefix code.
The different problems of synchronization have drawn the attention of many investigators (see for instance, surveys \cite{Ju},\cite{KV}, \cite{St}, \cite{Vo}).

A problem with a long story is the estimation of the minimal length of a synchronizing word.
  Jan \v{C}erny found in 1964 an $n$-state complete DFA with
shortest synchronizing word of length $(n-1)^2$ for an alphabet $\Sigma$ of size two \cite{Ce}.

\begin{con}
Any deterministic complete $n$-state synchronizing automaton over alphabet $\Sigma$ of letters on edges of the graph
$\Gamma$ has synchronizing word in $\Sigma$
of length at most $(n-1)^2$ \cite{Sta} (Starke, 1966).
  \end{con}

The cubic estimation on the upper bound of length of synchronizing word exists from 1966 [10].
The problem can be reduced to automata with a strongly connected graph \cite{Ce}.

We skip for brevity the luxuriant story of the conjecture with long list of references and several useful examples \cite{PS}, \cite{TM}.

The conjecture \v{C}erny holds true for a lot of automata, but in general
the problem still remains open although several hundreds of articles
consider this problem from different points of view \cite{TB}.
Moreover, two conferences "Workshop on Synchronizing  Automata" (Turku, 2004)
and  "Around the  \v{C}erny conjecture" (Wroclaw,2008) were dedicated to this longstanding conjecture.
The problem is discussed in "Wikipedia" - the popular Internet
Encyclopedia. Together with the Road
Coloring problem \cite{AW}, \cite{Fi}, \cite{T}, this simple-looking conjecture belongs to the most famous old combinatorial problems in the theory of finite automata \cite{KV}, \cite{MS}, \cite{PS}, \cite{St}, \cite{Vo}.

There are no examples of automata such that the length of
the shortest synchronizing word is greater than $(n-1)^2$.
Moreover, the examples of automata  with shortest synchronizing
word of length $(n-1)^2$ are infrequent. After the sequence found
by \v{C}erny and the example of \v{C}erny, Piricka and
Rosenauerova \cite{CPR}  of 1971 for $|\Sigma|=2$,
the next such examples were found by Kari \cite {Ka} in 2001 for $n=6$ and $|\Sigma|=2$ and by Roman \cite {Ro} for $n=5$ and
$|\Sigma|=3$ in 2004.

The package TESTAS \cite {Tt} founds five new automata with shortest synchronizing word of length $(n-1)^2$ for $n=3$ and
$n=4$ with small alphabet.

Don and Zantema present in \cite {DZ} an ingenious method of designing several new automata,
a kind of "hybrids" from existing examples of size three and four
from \cite{Ce}, \cite{CPR}, \cite {TS} and proved that
for $n\geq 5$ the method does not work.
So there are up to isomorphism exactly 15 DFA for $n=3$ and 12 DFA for $n=4$ with shortest synchronizing word of length $(n-1)^2$.

The authors of \cite {DZ} support the hypothesis from \cite{TS} (2006) that all automata with shortest synchronizing word of length $(n-1)^2$ are known,
of course, with essential correction found by themselves for
$n=3,4$.

Initially found upper bound for the minimal length of
synchronizing word was not even a polynomial \cite{Ce}.
For years, many authors have consistently improve this
estimation. The best known upper bound found by Frankl in 1982
\cite{Fr} is equal to $(n^3-n)/6$.
The result was reformulated in terms of synchronization in \cite{Pin} and repeated independently in \cite{KRS}.
The cubic estimation on the upper bound of length of reset word exists from 1966 \cite{Sta} and the estimation of Frankl was not changed since 1982.

 There are several reasons \cite{AGV}, \cite{TS} to believe that the length of the shortest synchronizing word for remaining automata is essentially less and the gap grows with $n$.

The considered deterministic automaton $A$ can be presented by a complete underlying graph with edges labelled by letters of an alphabet.

Our work uses a special class of matrices defined by words in
the alphabet of letters on edges of the underlying graph. Algebra with non-standard summation operation and its space of these matrices plays an impotent role in the study.
We study the rational series \cite{BR}. This approach for synchronizing  automata supposed first by B{\'e}al \cite{Be} proved to be fruitful \cite{CA}, \cite{Co}.

An effort to reduce the upper bound to $(n-1)^2$ and prove
the \v{C}erny conjecture is presented.

 \section*{Preliminaries}
We consider a complete $n$-state DFA with
 strongly connected underlying graph $\Gamma$
 over a fixed finite alphabet $\Sigma$ of labels on edges
of $\Gamma$ of an automaton $A$.

If there exists a path in an automaton from the state $\bf p$ to
the state $\bf q$ and the edges of the path are consecutively
labelled by $\sigma_1, ..., \sigma_k$, then for
$u=\sigma_1...\sigma_k \in \Sigma^+$ let us write
${\bf q}={\bf p}u$.

Let $Px$ be the subset of states ${\bf q}={\bf p}x$ for all
${\bf p}$ from the subset $P$ of states and $x \in \Sigma^+$.

 A word $s \in \Sigma^+ $ is called a {\it synchronizing} word of an automaton $A$ if $|As|=1$.
The word $s$ below denotes minimal synchronizing word such that for a state $\bf q$ $As=\bf q$.

The class of such automata is quite wide \cite{TI}.

The restriction on strongly connected graphs is based on \cite{Ce}.

The states $\bf p$ of the automaton $A$ are considered also as vertices of underlying graph $\Gamma$.

We connect a mapping of the set of states of the automaton
made by a word $u$ of $n\times n$-matrix $M_u$ such that for an element $m_{i,j} \in M_u$ takes place

\centerline{$m_{i,j}=1$ if the word $u$ maps ${\bf q}_i$ on ${\bf q}_j$ and $0$ otherwise.}

Any mapping of the set of states of the automaton  $A$ can be presented by a word $u \in \Sigma$
with corresponding matrix $M_u$.

 \centerline{$M_u=\left(
\begin{array}{ccccccc}
  0 & 0 & 1 & . & . & . &  0 \\
  1 & 0 & 0 & . & . & . &  0 \\
  0 & 0 & 0 & . & . & . &  1 \\
  . & . & . & . & . & . &  . \\
  0 & 1 & 0 & . & . & . &  0 \\
  1 & 0 & 0 & . & . & . &  0 \\
\end{array}\right)
$}

 Let us call the matrix $M_u$ of the mapping induced by the word $u$, for brevity, the matrix of word $u$, and vice versa,
the word $u$ also presents  the mapping of matrix $M_u$.
The matrices of word in arbitrary alphabet belong to a class of square matrices with one unit in every row and zeros in remaining cells. Let's call them also matrices of word.
Zero matrix also belongs to the considered algebra as zero and is denoted by zero. It is convenient to consider below zero matrix  as the matrix of the empty word.

Matrix multiplication in the algebra under consideration is standard. $M_uM_v=M_{uv}$ \cite{Be}.

The set of nonzero columns of $M_u$ (set of second indices of its elements) of $M_u$ is denoted as $R(u)$.

The subset of states $Au$ of the set of all states of $A$ is denoted $c_u$.

\section{The algebra of matrices of words and their properties}

\begin{rem} \label{r1}

Every unit in the product $M_uM_a$ is the product of two units, first unit from nonzero column of $M_u$
and second unit from a row of $M_a$.

The set $R(u)$ of nonzero columns of matrix $M_u$ corresponds the set of states $c_u$ of the automaton.
\end{rem}

\begin{lem}  \label {po}

The number of nonzero columns $|R(b)|$ is equal to the rank of $M_b$.

For the set of states of deterministic finite automaton and 
any words $u$ and $w$ $Auw \subseteq Aw$.

For every word $w$, $R(u)\subset R(v)$ implies
$R(uw)\subseteq R(vw)$.

${\bf p} \not\in Aw$ implies ${\bf p} \not\in Auw$ for the state ${\bf p}$.
 Nonzero columns of $M_{uw}$ have units also in $M_w$.

\end{lem}

Proof.
The matrix $M_b$ has submatrix with only one unit in every row and every nonzero column with nonzero determinant.
Therefore $|R(b)|$ is equal to the rank of $M_b$.

The properties of $Au \subseteq A$, $M_w$ and $M_{uw}$ follow from the definition of the matrix of word.

The set of nonzero columns of matrix defines a set of states.
The mapping by word $w$ of a set of states [columns] induces a mapping of its subset [columns].

For any word $u$ and the zero column of $M_w$
the corresponding column of $M_{uw}$ also consist of zeros.
Hence nonzero columns of $M_{uw}$ have units in $M_w$.

\begin{cor}  \label{c1}
The matrix $M_s$ of word $s$ is synchronizing if and only
if $M_s$ has zeros in all columns except one and units in
the residuary column. All matrices of right subwords of
$s$ also have at least one unit in this column.
\end{cor}

\begin{cor}  \label{c1a}
The invertible matrix $M_a$ keeps the number of units of any column of $M_u$ in corresponding column of the product 
$M_aM_u$.
\end{cor}

\begin{rem} \label{r4}

The columns of the matrix $M_uM_a$ are obtained by permutation of columns $M_u$.
Some columns can be merged (units of some columns are moved along
row to common column) with $|R(ua)|<|R(u)|$.

The rows of the matrix $M_aM_u$ are obtained by permutation of rows of the matrix $M_u$.
Some of these rows may disappear and replaced by another rows of $M_u$.

\end{rem}

\begin{lem} \label{l2}
For every words $a$ and $u$

\centerline{$|R(ua)| \leq |R(u)|$} and

\centerline{$R(au) \subseteq R(u)$.}
For invertible matrix $M_a$ $R(au)=R(u)$ and $|R(ua)|=|R(u)|$.

\end{lem}

Proof.
The matrix $M_a$ in the product $M_uM_a$ shifts column of
$M_u$ to columns of $M_uM_a$ without
changing the column itself (Remark \ref{r4}).
$M_a$ can merge columns of $M_u$.
In view of possible merged columns, $|R(ua)|\leq |R(u)|$.

Some rows of $M_u$ can be replaced in $M_aM_u$ by another row and therefore some rows from $M_u$ may be changed,
but zero columns of $M_u$ remain in $M_aM_u$ (Remark 2).
Hence $R(au) \subseteq R(u)$ and $|R(ua)| \leq |R(u)|$.

For invertible matrix $M_a$ in view of existence $M_a^{-1}$ we have $|R(ua)|=|R(u)|$ and $R(au)= R(u)$.

\subsection{Algebra with nonstandard summation of matrices
with one unit in every row}

\begin{lem}\label{lam}
 Suppose that for matrix $M_u$ of word $u$ and nonzero matrices $M_{u_i}$ of words $u_i$
\begin{equation}
M_u =\sum_{i=1}^k\lambda_i M_{u_i}. \label{lm}
\end{equation}
with coefficient $\lambda$ from some field.
Then the sum $\sum^k_{i=1}\lambda_i =1$ and the sum $S_j$ of values in every row $j$ of the sum in (\ref{lm}) also is equal to one.

If $\sum^k_{i=1}\lambda_iM_{u_i}=0$ (zero matrix) then
$\sum_{i=1}^k \lambda_i=0$ and $S_j=0$ for every $j$ with
$M_u=0$.

If the sum $\sum^k_{i=1}\lambda_i$ in every row is not unit
[zero] then $\sum_{i=1}^k\lambda_i M_{u_i}$
is not a matrix of word.
With respect to the considered operation, the set of matrices of words forms a space.

\end{lem}
Proof.
The nonzero matrices $M_{u_i}$ have $n$ cells with unit in the cell.
Therefore, the sum of values in all cells of the matrix $\lambda_i M_{u_i}$ is $n \lambda_i$.

For nonzero $M_u$ the sum is $n$. So one has in view of
$M_u =\sum_{i=1}^k\lambda_i M_{u_i}$

\centerline {$n=n\sum_{i=1}^k \lambda_i$, whence $1 =\sum_{i=1}^k \lambda_i$.}
Let us consider the row $j$ of matrix $M_{u_j}$ in (\ref{lm}) and let  $1_j$ be unit in the row $j$.
The sum of values in a row of the sum (\ref{lm}) is equal to unit in the row of $M_u$.
So $1 =\sum_{i=1}^k \lambda_i1_i=\sum_{i=1}^k \lambda_i$.

$\sum_{i=1}^k\lambda_i M_{u_i}=0$ implies $S_j=\sum_{i=1}^k \lambda_i1_i=\sum_{i=1}^k \lambda_i=0$ for  every row $j$.

If the matrix $M=\sum_{i=1}^k\lambda_i M_{u_i}$ is a matrix
of word or zero matrix then $\sum^k_{i=1}\lambda_i \in \{0, 1\}$.
If $\sum^k_{i=1}\lambda_i\not\in \{0, 1\}$ or
the sum is not the same in every row then we have opposite case and
the matrix does not belong  to the space of matrices of word.

\begin{rem} \label{r2}
The space generated by matrices of words has zero matrix of empty word and is closed under the summation in lemma above.
\end{rem}

\begin{lem}  \label {v3}
 The set $V$ of all $n\times k$-matrices of words
(or $n\times n$-matrices with zeros in fixed $n-k$ columns for $k<n$) has $n(k-1)+1$ linear independent matrices.
 \end{lem}

Proof.
Let us consider distinct $n\times k$-matrices of word with at most only one nonzero cell outside the last nonzero column $k$.

Let us begin from the matrices $V_{i,j}$ with unit in $(i,j)$ cell ($j<k$) and units in ($m,k$) cells for all $m$ except $i$.
The remaining cells contain zeros.
So we have $n-1$ units in the $k$-th column and only one unit in remaining $k-1$ columns of the matrix $V_{i,j}$.
Let the matrix $K$ have units in the $k$-th column and zeros in the other columns.
There are $n(k-1)$ matrices $V_{i,j}$. Together with $K$ they belong to the set $V$.
So we have $n(k-1)+1$ matrices. For instance,
\\
\\
\noindent$V_{1,1}=\left(
\begin{array}{cccccccc}
1 & 0 & 0 & . & 0 \\
0 & 0 & 0 & . & 1 \\
0 & 0 & 0 & . & 1 \\
. & . & . & . & . \\
0 & 0 & 0 & . & 1 \\
0 & 0 & 0 & . & 1 \\
\end{array}
\right)$
$V_{3,2}=\left(
\begin{array}{cccccccc}
0 & 0 & 0 & . & 1 \\
0 & 0 & 0 & . & 1 \\
0 & 1 & 0 & . & 0 \\
. & . & . & . & . \\
0 & 0 & 0 & . & 1 \\
0 & 0 & 0 & . & 1 \\
\end{array}
\right)$
$K=\left(
\begin{array}{cccccccc}
0 & 0 & 0 & . & 1 \\
0 & 0 & 0 & . & 1 \\
0 & 0 & 0 & . & 1 \\
. & . & . & . & . \\
0 & 0 & 0 & . & 1 \\
0 & 0 & 0 & . & 1 \\
\end{array}
\right)$

 The first step is to prove that the matrices $V_{i,j}$ and $K$ generate the space with the set $V$.
For arbitrary matrix $T$ of word from $V$ for every $t_{i,j} \neq 0$ and $j<k$,
let us consider the matrices $V_{i,j}$ with unit in the cell $(i,j)$ and the sum of them $\sum V_{i,j}=Z$.

The first $k-1$ columns of $T$ and $Z$ coincide.
   Hence in the first $k-1$ columns of the matrix $Z$ there is at most only one unit in any row.
 Therefore in the cell of $k$-th column of $Z$ one can find at most two values which differ by unit, say $m$ or $m-1$.
The value of $m$ appears if there are only zeros
in other cells of the considered row. Therefore $\sum V_{i,j} - (m-1)K=T$.

Thus, every matrix $T$ from the set $V$ is a span of
above-mentioned $(k-1)n +1$ matrices from $V$.
It remains now to prove that the set of matrices $V_{i,j}$ and $K$ is a set of linear independent matrices.

If one excludes a certain matrix $V_{i,j}$ from the set of these matrices, then it is impossible
to obtain a nonzero value in the cell $(i,j)$ and therefore to obtain the matrix $V_{i,j}$.
So the set of matrices $V_{i,j}$ is linear independent.
Every non-trivial span of the matrices $V_{i,j}$ equal to a matrix of word has at least one nonzero element
in the first $k-1$ columns.
Therefore, the matrix $K$ could not be obtained as a span of the matrices $V_{i,j}$.
Consequently the set of matrices $V_{i,j}$ and $K$ forms a basis of the set $V$.

\begin{cor}  \label {c2}
The set of all $n \times(n-1)$-matrices of words
(or $n\times n$-matrices with zeros in a fixed column)
has $(n-1)^2$ linear independent matrices.
 \end{cor}
Proof. For $k=n-1$ it follows from $n(n-1-1)+1= (n-1)^2$.

\begin{cor}  \label {c3}
There are at most $n(n-1)+1$ linear independent matrices of words in the set of $n\times n$-matrices.
 \end{cor}

 \begin{cor}  \label {c5}
There are at most $n+1$ linear independent matrices of words in the set of matrices with at most two nonzero columns.

There are at most $n$ linear independent matrices of words in the set of matrices with strictly two nonzero columns.

 \end{cor}

\begin{lem} \label{l3} {\it Distributivity from left.}

For every words $b$ and $x_i$

\centerline{$M_b\sum\tau_iM_{x_i}=\sum\tau_iM_bM_{x_i}$.}
\end{lem}

Proof.
The matrix $M_b$ shifts rows of every $M_{x_i}$ and of the sum of them in the same way according to Remark  \ref{r4}.
$M_b$ removes common row of them and replace also by common row (Remark \ref{r4}).

Therefore the matrices $M_bM_{x_i}$ has the origin rows of
$M_{x_i}$, maybe in another order,
and the rows of the sum $\sum\tau_iM_bM_{x_i}$ repeat rows of  $\sum\tau_iM_{x_i}$ also in the same order.

Note that this is not always true from right, the result is not necessary matrix of word.

\section{Rational series}
The section follows ideas and definitions from \cite{BR} and \cite{Be}.
We recall that a formal power series with coefficients in a field $K$ and variables in $\Sigma$ is
a mapping of the free monoid $\Sigma^*$ into $K$ \cite{BR}, \cite{CA}.

We consider an $n$-state automaton $A$. Let $P$ denote the subset
of states of the automaton with the characteristic column vector
$P^t$ of $P$ of length $n$ having units in coordinates corresponding to the states of $P$ and zeros everywhere else.
Let $C$ be a row of units of length $n$.
 Following \cite{Be}, we denote by $S$ the {\it rational series} depending on the set $P$ defined by:
\begin{equation}
(S,u) = CM_uP^t-CP^t= C(M_u-E)P^t. \label{ser}
\end{equation}

\begin{lem} \label{v4}
Let $S$ be a rational series depending on the
set $P$ of an automaton $A$.
Let $M_u=\sum_{j=1}^k\lambda_j M_{u_j}$.
Then $(S,u) =\sum_{j=1}^k\lambda_j (S,u_j)$.

If $(S,u_j)=i$ for every $j$ then also $(S,u)=i$.
\end{lem}

Proof.
One has in view of (\ref{ser})

\centerline {$(S,u)= C(\sum^k_{j=1}\lambda_jM_{u_j}-E)P^t$}
where $C$ is a row of units and $P^t$ is a characteristic
column of units and zeros.

Due to Lemma \ref{lam}

$\sum^k_{j=1}\lambda_jM_{u_j}-I=\sum^k_{j=1}\lambda_jM_{u_j}-\sum^k_{j=1}\lambda_j I =
\sum^k_{j=1} \lambda_j(M_{u_j}-I)$.

So
$(S,u)=C(M_u-I)P^t = C(\sum^k_{j=1}\lambda_j M_{u_j}-I)P^t =
\sum^k_{j=1}\lambda_jC(M_{u_j}-I)P^t=
\sum^k_{j=1}\lambda_j(S, u_j)$.

Thus, $(S,u) =\sum_{j=1}^k\lambda_j (S,u_j)$.

If $\forall j$ $(S,u_j)=i$, then
$(S,u)= \sum^k_{j=1}\lambda_j i=i\sum^k_{j=1}\lambda_j=i$ by Lemma \ref{lam}.

From Lemma \ref{v4} follows
\begin{cor} \label{c4}
Let $S$ be a rational series depending on the
set $P$ of an automaton.

The matrices $M_{u}$ with constant $(S,u)=i$ generate a space
$V$ of matrices of word such that for every nonzero matrix $M_t \in V$ of word $t$ $(S,t)=i$.

The space $V$ with the summation operation, presented in previous lemmas, possesses many natural concepts such as
linear independence,  basis and dimension.

\end{cor}

\section{The equation with unknown matrix $L_x$}

Remember that $As=\bf q$ for minimal synchronizing word $s$.
Let the state $q$ have number one.

\begin{defn} \label{dL}
Let $S_q$ be a rational series depending on the set
$P =\{\bf q\}$ of size one of nonzero column $q$ of $M_s$.

If the set of cells with units in the column $\bf q$ of
$M_x$ and $M_y$ are equal then

 \centerline{$M_x \sim_q M_y$,}

if this set of $M_x$ is a subset of the analogous set of $M_y$ then we write

\centerline{$M_x \sqsubseteq_q M_y$.}

\end{defn}
The solution $L_x$ of the equation
\begin{equation}
M_uL_x=M_s \label{ux}
\end{equation}
for synchronizing matrix $M_s$ and arbitrary $M_u$ with words $u, s \in \Sigma$ and $As=\bf q$ must have units in the column of the
state $\bf q$ and have one unit in every row with rest of zeros as a matrix of word.
(See Lemmas \ref{lam}, \ref{v3} and \ref{l3}
about algebraic properties of algebra of matrices of word.)

In general, there are some solutions $L_x$ of synchronizing
continuations $x$ of the word $u$ in synchronizing word.

\begin{lem}  \label{l5}
Every equation $M_uL_x=M_s$ (\ref{ux}) has a solutions $L_x$
with $(S_q,x) \geq 0$.

 $|R(u)|-1=(S_q,x)$ for $L_x$ with minimal $(S_q,x)$ (a minimal solution).

Let every matrix $L_y$ satisfies the equation (\ref{ux}) iff
$L_x\sqsubseteq_q L_y$.

There exists one-to-one correspondence between nonzero columns of $M_u$, units in the column $q$ of minimal solution $L_x$
and the set of states $c_u=Au$ of automaton $A$.

The rank $|R(x)|\leq n-(S_q,x)$ and $R(u)+R(x)\leq 1$. The equality is possible if
no more than one unit per columns exists, except $q$.

\end{lem}

Proof.
The matrix $M_s$ of rank one has column $q$ of units of the state $\bf q$  ($As=\bf q$).

For every nonzero column $j$ of $M_u$ with elements
$u_{i,j}=1$ and $s_{i,q}=1$ in the matrix $M_s$ let the cell $(j,q)$ have unit in the matrix $L_x$.
So the unit in the column $q$ of matrix $M_s$ is a product of
every unit from the column $j$ of $M_u$
and unit in the cell $(j,q)$ of column $q$ of $L_x$, whence
$(S_q,x) \geq 0$.

The set $R(u)$ of nonzero columns of $M_u$ corresponds the set of cells of the column $q$ with unit of minimal $L_x$.
$(S_q,u)=CM_uP^t-CP^t$ (\ref{ser}) for $CP^t=1$ and $CM_uP^t=|R(u)|$. So $(S_q,x)=|R(u)|-1$
and the minimal solution $L_x$ has $(S_q,x)+1$ units in the column $q$.

So the column $q$ of every solution has at least $|R(u)|-1$ units.

Units in rows of $L_x$ corresponding zero columns of $M_u$ do not imply on result in (\ref{ux}) (Remark \ref{r1}
and therefore can be placed arbitrarily, of course, one unit in a row.
The remaining empty cells obtain zero in matrices of word
$L_ x$.

Lastly every solution $L_x$ of (\ref{ux}) has one unit with rest of zeros in every row and can be called a matrix of word.

Zeros in the cells of column $q$ of minimal $L_x$ correspond zero columns of $M_u$.
Therefore for the matrix $L_y$ such that
$L_x \sqsubseteq_q L_y$ we have $M_uL_y=M_s$.
Every solution $L_y$ must have units in cells of column $q$ that correspond $|R(u)|=(S_q,x)+1$ nonzero columns of $M_u$
and minimal $L_x$.

Thus, the equality $M_uL_x=M_uL_y=M_s$ is
equivalent to $L_x \sqsubseteq_q L_y$ for the minimal $L_x$.
The set $R(x)$ in (\ref{ux}) has therefore at most
$n-(S_q,x)-1=n-R(u)$ nonzero columns besides $q$, whence the rank $|R(x)|\leq n-(S_q,x)$ and $R(u)+R(x)\leq 1$.

The equality $|R(x)|=n-(S_q,x)$ and $R(u)+R(x)=1$ is possible when all these $n-(S_q,x)-1$ columns besides $q$ are columns with one unit.

The matrix $M_u$ with set $R(u)$ of nonzero columns maps
the automaton on the set $c_u$ of states and on the set of units in the column $q$ of minimal $L_x$.

\begin{cor} \label{c7}

For minimal solution $L_x$ of the equation
$M_uL_x=M_s$ and minimal solution $L_y$ of the equation
$M_{ut}$ for every $t$ $L_y=M_s$ one has 
$(S_q,y)\leq (S_q,x)$.

One can reduce $(S_q,x)$ of minimal solution $L_x$ in
(\ref {ux}) and extend the rank $R(x)$ only by
decrease of $R(u)$.

The matrices $L_x$ with $(S_q,x)\geq d>0$ for integer $d$  generate a space of dimension at most
$n(n-d-1) \leq n(n-2)$.

\end{cor}

 The proof follows from Lemma \ref{l2} in view of
$|R(ut)|\leq |R(u)|$ and Lemma \ref{v3}.

\subsection{Right pseudoinverse matrices}

\begin{defn} \label{inv}

Let us call the matrix $M_{a^-}$ of word $a^-$
{\sf right pseudoinverse} matrix of the matrix $M_a$
of a word $a$ if for precisely one element $a_{i,j}=1$ of
every nonzero column $j$ of $M_a$ the cell $(j,i)$ of
$M_{a^-}$ has unit.

In still zero rows of $M_{a^-}$ is added one unit arbitrarily in every such row.
Zeros fill rest of cells. So it is a matrix of word.

\end{defn}

For instance,
\\
\\
\noindent$M_a=\left(
\begin{array}{cccccccc}
  0 & 1 & 0 & 0 & 0 \\
  0 & 1 & 0 & 0 & 0 \\
  0 & 0 & 0 & 0 & 1 \\
  0 & 0 & 1 & 0 & 0 \\
  0 & 0 & 1 & 0 & 0 \\
\end{array}
\right)$
$M_{a^-}=\left(
\begin{array}{cccccccc}
  0 & 1 & 0 & 0 & 0 \\
  \d{1} & 0 & 0 & 0 & 0 \\
  0 & 0 & 0 & 0 & \d{1} \\
  0 & 0 & 0 & 1 & 0 \\
  0 & 0 & \d{1} & 0 & 0 \\
\end{array}
\right)$
$M_{a^-}=\left(
\begin{array}{cccccccc}
  1 & 0 & 0 & 0 & 0 \\
  0 & \d{1} & 0 & 0 & 0 \\
  0 & 0 & 0 & \d{1} & 0 \\
  0 & 0 & 0 & 0 & 1 \\
  0 & 0 & \d{1} & 0 & 0 \\
\end{array}
\right)$

\begin{rem} \label{ri}
For invertible matrix $M_a$ (with $|R(a)|=n$) we have
a special case $M_{a^-}=M_a^{-1}$, for singular $M_a$ there are some pseudoinverse matrices, even some invertible.

The product $M_aM_{a^-}$ does not depend on arbitrary adding of units in rows of $M_{a^-}$ corresponding
zero columns of $M_a$ in view of Remark \ref{r1}, since the product needs at least one nonzero cell in the column for
singular $M_a$.
\end{rem}

\begin{rem} \label{rm}
Some matrix $M_{s^-}$ with $As=q$ defines paths of $s$ from the state $q$ in opposite direction to every state.

Some matrix $M_{a^-}$ defines several paths of $a$ from the
state $q$ in opposite direction to states corresponding
states $c_a$ and nonzero columns of $M_a$.

Thus, pseudoinverse matrices can be considered as matrices of word in the alphabet
$\Sigma^-$ (in their first part defined by generic matrix).
\end{rem}

For $M_{a^-}=\sum \lambda_iM_{a_i^-}$, pseudoinverse $M_{b^-}$ and $M_{a_i^-}$ by Lemma \ref{l3}

\centerline{$M_{b^-}M_{a^-}=\sum \lambda_iM_{b^-}M_{a_i^-}$.}

\begin{lem} \label{l7}
For every equation $M_uL_x=M_s$ and every letter $\beta$ the equation

\begin{equation}
 M_{u\beta}L_y=M_s      \label{yx}
\end{equation}
has solution $L_y$. For minimal solutions $L_x$ of (\ref{ux}) and
$L_y$ one has $(S_q,y) \leq (S_q,x)$. $R(y)\supseteq R(x)$ is
possible.

For every solution $L_x$ of equation (\ref{ux}) and
suitable $M_{\beta^-}$, even invertible,

\centerline{$M_s=M_uM_{\beta}M_{\beta^-}L_x$}
for solution $L_y=M_{\beta^-}L_x$ of the equation (\ref{yx}).

Let $|R(u)|=|R(u\beta)|$.
Then $(S_q,y)=(S_q,x)$ for minimal solutions
$L_y$, $L_x$ and  for invertible $M_{\beta^-}$ maximal ranks
$R(y)=R(x)$.
But for invertible $M_{\beta^-}$ anyway $R(y)=R(x)$.

For $|R(u)|\neq |R(u\beta)|$ and singular $M_{\beta^-}$
there exists solution $L_y$ of the equation
$M_uM_{\beta}L_y=M_{u\beta}L_y=M_s$
such that $(S_q,y)<(S_q,x)$ for minimal solutions and
$|R(y)|>|R(x)|$ for maximal ranks.
Thus for some $L_y$ one has $R(x) \subset R(y)$ and
$|R(y)|=|R(x)|+|R(u)|-|R(u\beta)|$.
 $|R(u)|+|R(x)|=n+1$ for last considered word $u$ and corresponding minimal solution $L_x$.

  \end{lem}

Proof.
The equality in (\ref{yx}) is correct for some $L_y$.
By Lemma \ref{l2} $|R(u)|\geq |R(u\beta)|$.
Therefore by Corollary \ref{c7}
$(S_q,y) \leq (S_q,x)$ for minimal solutions $L_x$ and $L_y$.
Hence in view of arbitrary placing $n-(S_q,y)$ units
in $L_y$ outside column $q$ (Lemma \ref{l5}),
$R(x) \subseteq R(y)$ is possible for some minimal $L_y$.

The matrix $M_{\beta^-}$ returns the set of nonzero columns
from $R(u\beta)$ to $R(u)$
(or to its part) in view of Definition \ref{inv}.

Arbitrary placing of units in some rows of $M_{\beta^-}$ does not change the product $M_{\beta}M_{\beta^-}$
by Remark \ref{ri}.
Therefore $|R(u)M_{\beta}M_{\beta^-}|\leq |R(u)|$,
whence $(S,x)\geq(S,y)$.
Hence the equality in

\centerline{$M_uM_{\beta}M_{\beta^-}L_x=M_{u\beta}M_{\beta^-}L_x=M_{u\beta}L_y=M_s$}
is correct for some $L_y=M_{\beta^-}L_x$ with
$R(x)\subseteq R(y)$ and free placing only of
$(S_q,x)-(S_q,y)$ units in $L_y$ (see Lemma \ref{l5}).

In the case $|R(u)|=|R(u\beta)|$ the matrix $M_{\beta}$
does not merge some columns of $M_u$ and by Lemma \ref{l5}
$(S_q,y)=(S_q,x)$ for minimal solutions $L_y$ and $L_x$.
So $R(y)=R(x)$ in view of $R(x)\subseteq R(y)$ and
Lemma \ref{l5}.
For invertible matrix $M_{\beta^-}$ also
$R(y)=M_{\beta^-}R(x)=R(x)$.

From $|R(u)|\neq |R(u\beta)|$ due to Lemma \ref{l2} follows
$|R(u\beta)|<|R(u)|$, whence for some solution $L_y$
of the equation $M_uM_{\beta}L_y=M_s$
$(S_q,y)<(S_q,x)$ for both such minimal solutions by Lemma
\ref{l5}.

 After filling by units nonzero columns of $R(x)$ by units from $R(y)$,
$R(x)$ can be extended by new columns using arbitrary addition
of $|R(y)|-|R(x)|$ units and $R(y)\supset R(x)$.

The possible equalities $|R(x)|=n-(S_q,x)$ and
 $|R(u)|-1=(S_q,x)$ (Lemma \ref{l5}) imply
 for minimal $L_y$ and $L_x$ and maximal ranks
$|R(u)|+|R(x)|=n+1$,
$|R(y)|-|R(x)|=(S_q,x)-(S_q,y)$ and
$|R(y)|=n-(S_q,y)=|R(x)|+(S_q,x)-(S_q,y)=
|R(x)|+|R(u)|-|R(u\beta)|$.

From Lemma \ref{l7} follows

\begin{cor}  \label{c9}
A set of linear independent solutions $L_x$ of
(\ref{ux}) and $L_x$ with constant rational series $(S_q,x)$
and fixed $R(x)$
 can be expanded sometimes by help of invertible matrices
$M_{\beta^-}$ of letters $\beta^-$ in the alphabet
$\Sigma^-$ (and words of them)
with the same series $(S_q,x)$ and common set $R(x)$.
\end{cor}
Proof.
The invertible matrix $M_{\beta^-}$ does not change $(S_q,x)$ and $R(x)$ of matrix $L_x$ in the equation (\ref{yx}) by
Corollary \ref{c1a}.

Anyway we have a space generated by matrices with one unit in every row and with rest of zeros.

\begin{rem} \label{r9}
Not minimal solutions $L_y$ of (\ref{yx}) with
$(S_q,y)>(S_q,x)$ and $R(y) \subset R(x)$ also are useful sometimes for extending subspace $V_k$ of greater $(S_q,y)$.
Arbitrary placing of units in $L_y$ is preferable in nonzero columns of matrices of $V_k$.
\end{rem}

\begin{lem} \label{v8}

Let the space $W$ be generated by solutions $L_x$
of the equation $M_uL_x=M_s$ 
of words $u$ of length at most $k$. Let $L_x$ have
common zero column.

Then there exist a word $u$ of length at most $k+1$
such that the solution $L_x \not\in W$.

\end{lem}

 Proof. Assume the contrary: for every word $u$ with
$|u|\leq k$ and every letter $\beta$
the equation $M_{u\beta}L_y=M_s$ has every solution
$L_y\in W$ for every word $u\beta$ of length at most
$k+1$. The matrix $M_{\beta^-}L_x$ has units only in
nonzero columns of $L_x$ by Corollary \ref{c1a}.

Therefore every solution $L_y=M_{t^-}L_x$ of the equation 
$M_{u\beta}L_y=M_s$ belongs to $W$ by induction 
for arbitrary word $t$. (Lemma \ref{l7}).
The matrix $L_y$ has units only in nonzero columns 
of matrices $W$.

The considered automaton is synchronizing, whence there 
exists $|R(u)|=1$. 
By Lemma \ref{l5} $|R(u)|-1=(S_q,x)$, whence
$(S_q,x)=0$ for some solution $L_x$ of (\ref{ux}).

Therefore the existence of common zero column in all 
solutions $L_x$ of (\ref{ux}) implies contradiction.

\section{The sequence of words of growing length}
The space $W_j$ is generated by $M_s$ and
linear independent solutions $L_x$ of equations
 $M_uL_x=M_s$ (\ref{ux}) with $|u|\leq j$.
Let's use induction below.

The space $W_0$, in particular, is generated by minimal synchronizing matrix $M_s$,
a trivial solution of every equation (\ref{ux}).
$\dim(W_0)=1$.
The minimal solution $L_x$ of equation $M_{\alpha}L_x=M_s$
for the left letter $\alpha$ of $s$ and $M_s$
generate the subspace $W_1$. $\dim(W_1)=2$.

We consider for every $W_j$ the set of solutions $L_x$ of equation (\ref{ux}) for $|u|\leq j+1$.
We choose a solution $L_x \not\in W_j$ for minimal such
$|u|$ following Lemmas \ref{l5}, \ref{l7}.
The possibility of existence such $L_x$ is studied
in Lemma \ref{v8}.
Then $L_x$ is added to the space $W_j$ turning it into
the space $W_{j+1}$ with corresponding growth of $j$.

The solutions $L_x$ of equations $M_uL_x=M_s$ with fixed $(S_q,x)=n-i$ generate subspace $V_i\subseteq W_j$ with
the same $(S_q,x)$ by Corollary \ref{c4}.
$V_i$ can be extended by help of invertible matrices
$M_{\beta^-}$ of letters with keeping by Corollary \ref{c5} the same rank $|R(x)|$.
In view of Lemma \ref{l7} the set of nonzero columns in
matrices $V_i$ does not changed because $M_{\beta^-}$
is invertible.

We can extend $V_i$ following Corollary \ref{c9} or reduce $(S_q,x)=i$.

The space $W_j$ is created by generators $L_x$ of subspaces $V_i$ by decrease of $i$ from $i=n-1$ until $i=1$.
So $(S_q,x)>0$ for every generator $L_x$ of $W_j$.

With decreasing of $(S_q,x)$ and increasing $R(x)$, we can add to the set of nonzero columns of the set of matrices $L_x$ new
columns due to $R(y)\supset R(x)$ and
$|R(y)=|R(x)|+(S_q,x)-(S_q,y)$ (Lemma \ref{l7}).

The set of nonzero columns of $W_j$ is a union of  nonzero columns of $V_i$ and in view of $(S_q,x)>0$
there is common zero column in matrices $W_j$.

One can extend the rank $R(x)$ and reduce $(S_q,x)$ of minimal solution $L_x$ of (\ref{ux}) only by decreasing $R(u)$
(Lemma \ref{l7}, Corollary \ref{c7}).

We follow conditions of Lemma \ref{l7} with a view to obtain $(S_q,x)=0$ for solution $L_x$ of (\ref{ux}).
It's unavoidable after $|dim(W_j)>n(n-2)$ (or before) in view of Corollary \ref{c2} and Lemma \ref{v8}.

The distinct linear independent solutions can be added consistently
extending the dimension of $W_j$ and upper bound $j$
of the length of the word $u$. So
\begin{equation}
\dim(W_j)=j+1 \quad |u|\leq j.     \label{d}
\end{equation}

\section{Theorems}

\begin{thm} \label{t}

The deterministic complete $n$-state synchronizing automaton
$A$ with strongly connected underlying graph over alphabet $\Sigma$ has synchronizing word in $\Sigma$
of length at most $(n-1)^2$.

\end{thm}

Proof.
The introduction to the former section considers a growing sequence of spaces $W_j$
(an ascending chain by dimension $j+1$) generated by linear
independent solutions $L_x$ of the equations (\ref{ux})
for $|u|\leq j$ by help of Lemmas \ref{l5} and \ref{l7} with Corollaries.

By Lemma \ref{v8}, any space $W$ generated by solutions $L_x$ of the equation $M_uL_x=M_s$ with common zero column of restricted
length $|u|\leq k$ has a solution $L_x \not\in W$ for some word $u$ of length at most $k+1$.

 $\dim(W_j)\leq n(n-2)+1$ for $W_j$ with matrices
having units in at most $n-1$ column by Corollary \ref{c2} 
of Lemma \ref{v3}.

So at least one solution $L_y\not\in W$ of equation (\ref{yx}) has corresponding word $v$ with $|v|=j+1$
and minimal $(S_q,y)=0$.

By Lemma \ref{l5}, for $L_x$ with minimal $(S_q,x)$ of equation (\ref{ux}) $|R(u)|-1=(S_q,x)$.
We reach finally a minimal $(S_q,y)=0$
for path of length $|v| \leq n(n-2)+1$.
So

\centerline{$|v|\leq n(n-2)+1$ with $|R(v)|=1$ and
$(S_q,y)=0$.}
Consequently the matrix $M_{v}$ of rank one in equation $M_vL_y=M_s$ is the matrix
of synchronizing word $v$ of length at most
$n(n-2)+1=(n-1)^2$.
\\
\\
In view of Lemma \ref{v3} with Corollaries from Theorem
\ref{t} follows

\begin{cor} \label{c14}
For every integer $k<n$ of deterministic complete $n$-state synchronizing automaton $A$ with strongly connected
underlying graph over alphabet $\Sigma$
there exists a word $v$ of length at most $n(k-1)+1$ such that $|Av|\leq n-k$.
\end{cor}

\begin{thm}\label{t2}
The deterministic complete $n$-state synchronizing automaton
$A$ with underlying graph
over alphabet $\Sigma$ has synchronizing word in $\Sigma$ of length at most $(n-1)^2$.
\end{thm}
Follows from Theorem \ref{t} because the restriction for strongly connected graphs can be omitted due to \cite{Ce}.

\begin{thm}\label{t4}
Suppose that $|\Gamma\alpha|<|\Gamma|-1$ for a letter $\alpha \in\Sigma$ in deterministic complete $n$-state
synchronizing automaton $A$ with underlying graph $\Gamma$ over alphabet $\Sigma$.

Then the minimal length of synchronizing word of the automaton is less than $(n-1)^2$.

\end{thm}
Proof.
We follow the proof of Theorem  \ref{t}.

The difference is that at the beginning of the proof the equation
(\ref{ux}) has at least two linear independent nontrivial solutions
for the matrix $M_{\alpha}$ of a letter $\alpha$ equal to the first word $u$ of length one.

Hence we obtain finally synchronizing word of length less
than $(n-1)^2$.

Let us go to the case of not strongly connected underlying graph with $n-|I|>0$ states
outside minimal strongly connected ideal $I$.

This ideal has synchronizing word of length at most $(|I|-1)^2$ (Theorem \ref{t}).
There is a word $p$ of length at most $(n-|I|)(n-|I|+1)/2$ such that $Ap \subset I$.

$(|I|-1)^2 +(n-|I|)((n-|I|)+1)/2<(n-1)^2$.
Thus, the restriction for strongly connected automata can be omitted.

\begin{thm}\label{t5}
Every road coloring of edges of $n-state$ strongly connected directed graph
with constant outdegree and gcd=1 of length of all its cycles has synchronizing word
of length at most $(n-1)^2$.
\end{thm}

Proof follows from Theorem \ref{t} and work \cite{T}.

\section*{Examples of synchronizing word of length $(n-1)^2$.}

Units in vector of $c_u$ correspond nonzero columns from
$R(u)$ of matrix $M_u$.
The vector of $c_u$ is equal to column $q$ of solution $L_x$ of equation $M_uL_x=M_s$ (Lemma 7).

The matrices $L_x$ corresponding word $u$ of $M_u$ (or $L_v$ where $L_x\sqsubseteq_q L_v$) of fixed $(S_g,x)$
are linear independent in lines of examples below.

The words $u$ below are ordered according the length $|u|$.
Lemma \ref{l7} was used together with Remark \ref{r9}.
$|R(u)|+|R(x)|=n+1$ (or less in view of Remark \ref{r9}).

J. Kari \cite{Ka} discovered the following example of $6$-state automaton with minimal synchronizing word of length $(n-1)^2$.

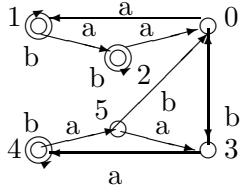
\begin{picture}(130,70)
 \end{picture}
\begin{picture}(130,74)
\multiput(6,60)(64,0){2}{\circle{6}}
\multiput(6,13)(64,0){2}{\circle{6}}
 \multiput(22,56)(22,0){2}{a}
\multiput(16,19)(34,0){2}{a}
 \put(36,21){\circle{6}}
\put(36,48){\circle{6}}
 \put(7,14){\vector(4,1){28}}
\put(7,57){\vector(4,-1){26}}

\put(39,52){\vector(4,1){27}}
 \put(37,20){\vector(4,-1){28}}
\put(67,63){\vector(-1,0){57}}
 \put(36,64){a}
\put(67,12){\vector(-1,0){57}}
 \put(32,0){a}

\put(70,15){\vector(0,1){42}}
 \put(70,59){\vector(0,-1){42}}
\put(34,21){\vector(1,1){36}}
 \put(52,28){b}

  \put(76,22){b}
\put(76,10){3}
\put(76,60){0}
\put(27,25){5}
\put(43,38){2}

\put(25,37){b}
\put(36,48){\circle{10}}
\put(-6,10){4}
\put(0,20){b}
\put(-6,60){1}
\put(0,45){b}
\put(37,42){\vector(2,1){4}}

\put(6,60){\circle{10}}
\put(4,64){\vector(2,1){4}}

 \put(6,13){\circle{10}}
\put(7,7){\vector(2,1){4}}
 \end{picture}

The minimal synchronizing word

\centerline{$s=\it ba^2bababa^2b^2aba^2ba^2baba^2b$}
has the length at the \v{C}erny border.

Every line below presents a pair (word $u$, $n$-vector $c_u$) of the word $u$.

$(b, 111110)$ $|R(u)|=5$, $|R(x)|=2$

$(ba, 111011)$

$(ba^2, 111101)$

$(ba^2b, 111100)$ $|R(u)|=4$

$(ba^2ba, 111010)$

$(ba^2bab, 011110)$

$(ba^2baba, 101111)$ $|R(v)|=5$ (l01011 of $|R(u)|<|R(v)|$)

$(ba^2babab, 101110)$ $|R(u)|=4$

$(ba^2bababa, 110101)$

$(ba^2bababa^2, 011101)$

$(ba^2bababa^2b, 111000)$ $|R(u)|=3$

$(ba^2bababa^2b^2, 011100)$

$(ba^2bababa^2b^2a, 110111)$ $|R(v)|=5$ (101010 of $|R(u)|<|R(v)|$)

$(ba^2bababa^2b^2ab, 001110)$ $|R(u)|=3$

$(ba^2bababa^2b^2aba, 100011)$

$(ba^2bababa^2b^2aba^2, 011111)$  $|R(v)|=5$ (010101 of
$|R(u)|<|R(v)|$)

$(ba^2bababa^2b^2aba^2b, 110000)$ $|R(u)|=2$

$(ba^2bababa^2b^2aba^2ba, 011000)$

$(ba^2bababa^2b^2aba^2ba^2, 101000)$

$(ba^2bababa^2b^2aba^2ba^2b, 001101)$ $|R(v)|=3$ (001100 of
$|R(u)|<|R(v)|$)

$(ba^2bababa^2b^2aba^2ba^2ba, 100010)$ $|R(u)|=2$

$(ba^2bababa^2b^2aba^2ba^2bab, 000110)$

$(ba^2bababa^2b^2aba^2ba^2baba, 001011)$  $|R(v)|=3$ (000011 of $|R(u)|<|R(v)|$)

$(ba^2bababa^2b^2aba^2ba^2baba^2, 000101)$ $|R(u)|=2$

$(ba^2bababa^2b^2aba^2ba^2baba^2b=s, 100000)$  $|R(s)|=1$

By the bye, the matrices of right subwords of $s$ are simply linear independent.

This property is by no means rare for minimal synchronizing word.
\\
\\
For the \v{C}erny sequence of $n$-state automata \cite{Ce} the situation is more pure.

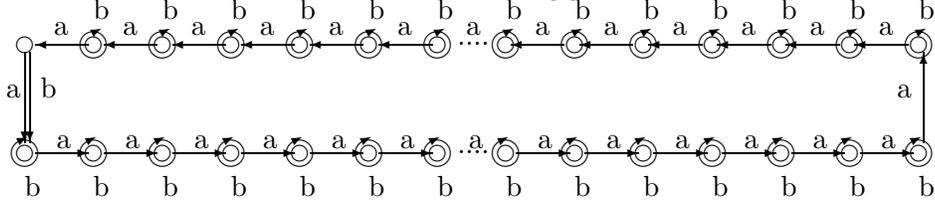
\begin{picture}(300,70)
\multiput(0,54)(26,0){14}{\circle{6}}
\multiput(26,54)(26,0){13}{\circle{10}}
\multiput(24,54)(26,0){6}{\vector(-1,0){20}}
\multiput(204,54)(26,0){6}{\vector(-1,0){20}}
\put(160,54){ ....}

\multiput(11,58)(26,0){13}{a}
 \multiput(26,64)(26,0){13}{b}
 \multiput(25,58)(26,0){13}{\vector(2,1){4}}
\put(340,17){\vector(0,1){34}}
 \put(0,51){\vector(0,-1){34}}
\put(2,51){\vector(0,-1){34}}
\put(-7,34){a} \put(6,34){b} \put(330,34){a}
\multiput(0,13)(26,0){14}{\circle{6}}
\multiput(0,13)(26,0){14}{\circle{10}}
\multiput(4,13)(26,0){6}{\vector(1,0){20}}
\multiput(186,13)(26,0){6}{\vector(1,0){20}}
\put(160,13){ ....}

\multiput(12,15)(26,0){13}{a}
 \multiput(0,-3)(26,0){14}{b}
 \multiput(-3,17)(26,0){14}{\vector(2,1){4}}
 \end{picture}
\\
\\
The minimal synchronizing word

\centerline{$s=b(a^{n-1}b)^{n-2}$}
 of the automaton also has the length at the \v{C}erny border. For $n=4$
\\
\\
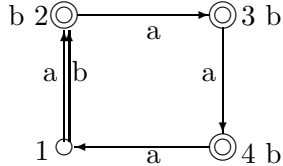
\begin{picture}(140,57)
\end{picture}
\begin{picture}(140,57)
\multiput(-21,60)(60,0){2}{\circle{6}}
\multiput(-21,60)(60,0){2}{\circle{10}}

 \multiput(-21,10)(60,0){2}{\circle{6}}

\put(39,10){\circle{10}}

 \put(-21,12){\vector(0,1){42}}
\put(-19,12){\vector(0,1){42}}

 \put(34,10){\vector(-1,0){51}}
\put(-16,60){\vector(1,0){50}}
 \put(39,55){\vector(0,-1){40}}

  \multiput(-29,35)(60,0){2}{a}
\multiput(10,4)(0,47){2}{a}

\put(-42,56){b 2}
 \put(-32,5){1}
\put(46,56){3 b}
 \put(46,4){4 b}
\put(-18,35){b}

 \end{picture}

and synchronizing word  $baaabaaab$ with pairs of word $u$ and
$n$-vector of $c_u$ of linear independent matrices $L_u$ below.

$(b, 0111)$ $|R(u)|=3$

$(ba, 1011)$

$(baa, 1101)$

$(baaa, 1110)$

$(baaba, 1010)$  $|R(u)|=2$

$(baaaba, 0011)$

$(baaabaa, 1001)$

$(baaabaaa, 1100)$ $|u|=8$

$(baaabaaab=s, 0100)$ $|R(s)|=1$
\\
\\
In the example of Roman \cite{Ro}

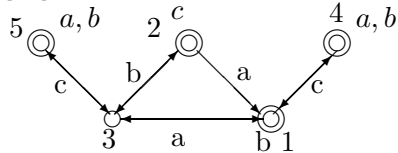
\begin{picture}(140,54)
\end{picture}
\begin{picture}(140,50)
\multiput(-21,39)(56,0){3}{\circle{6}}
\multiput(-21,39)(56,0){3}{\circle{10}}

 \multiput(6,10)(60,0){2}{\circle{6}}

\put(66,10){\circle{10}}
 \put(5,12){\vector(-1,1){24}}
  \put(-19,36){\vector(1,-1){24}}

\put(67,12){\vector(1,1){22}}
  \put(91,36){\vector(-1,-1){22}}

  \multiput(-16,20)(97,0){2}{c}

\put(7,12){\vector(1,1){24}}
\put(31,36){\vector(-1,-1){24}}

\put(38,36){\vector(1,-1){24}}

   \put(9,10){\vector(1,0){54}}
   \put(63,10){\vector(-1,0){54}}

\put(2,-1){3}
 \put(28,0){a}
\put(53,25){a}
 \put(11,24){b}

\put(-33,41){$5$}
\put(-14,45){$a,b$}
\put(28,48){$c$}
\put(19,41){$2$}
\put(98,45){$a,b$}
\put(88,47){$4$}
\put(60,-2){b   1}
 \end{picture}

the minimal synchronizing word

\centerline{$s=ab(ca)^2c$ $bca^2c$ $abca$}

The line below presents a pair (word $u$, $n$-vector of $c_u$) of the word $u$.

$(a, 10111)$ $|R(u)|=4$

$(ab, 11011)$

$(abc, 11110)$

$(abca, 10110)$ $|R(u)|=3$

$(abcac, 10011)$

$(abcaca, 01111)$ $|R(v)|=4$  (00111 of  $|R(u)|<|R(v)|$|)

$(abcacac, 10101)$ $|R(u)|=3$

$(abcacacb, 11001)$

$(abcacacbc, 01110)$

$(abcacacbca, 10010)$ $|R(u)|=2$

$(abcacacbca^2, 00110)$

$(abcacacbca^2c, 10001)$

$(abcacacbca^2ca, 11101)$ $|R(v)|=4$ (00101 of $|R(u)|<|R(v)|$)

$(abcacacbca^2cab, 01001)$   $|R(u)|=2$

$(abcacacbca^2cabc, 01100)$

$(abcacacbca^2cabca=s, 10000)$ $|R(s)|=1$

\section*{Acknowledgments}
I would like to express my gratitude to Francois Gonze,
Dominique Perrin, Marie B{\'e}al, Akihiro Munemasa,
Wit Forys as well as to Benjamin Weiss, Mikhail Volkov, Mikhail Berlinkov and Evgeny Kleiman for fruitful and essential remarks throughout the study.

 \end{document}